\documentclass[aps,twocolumn,prl]{revtex4}
\usepackage{graphicx}
\usepackage{epsfig}
\begin{document}
\title{Projecting Fermion Pair Condensates into Molecular Condensates}
\author{Roberto B. Diener and Tin-Lun Ho}
\address{Department of Physics,  The Ohio State University,
Columbus, Ohio
43210}

\begin{abstract}
We offer strong evidence that the recent observations by M. Greiner, C. Regal, and D. Jin\cite{Jin} and by MIT group\cite{Ketterle}  are signatures of a fermion superfluid in the strongly interacting regime made up of large fermion pairs. Our conclusions are based on calculations using crossover theory for different potentials including those with the characteristics of two-channel models.  Our results demonstrate clearly universality near resonance. The $T_{c}$ predicted by crossover theory is a perfect match with the observed boundary of vanishing condensate fraction with no adjustable parameters.  
\end{abstract}

\maketitle

In an exciting recent paper, M. Greiner, C. Regal, and D. Jin reported evidence for condensation of fermion pairs of $^{40}$K near Feshbach resonance\cite{Jin}. Similar results in $^{6}$Li  were reported later by Ketterle's group at MIT but with a different interpretation\cite{Ketterle}.  Feshbach resonance is produced by Zeeman shifting the energy of a bound state in closed channel to zero energy. When this energy (linear in magnetic field $B$ and usually referred to as ``detuning" $\delta$) vanishes, a low energy fermion pair in the open channel 
strongly resonates with the bound state  through hyperfine interaction, leading to considerable scattering. 
The regions $\delta>0$ and $\delta <0$ (which has negative and positive scattering length $a_{s}$ respectively)
are referred to the BCS and BEC side of the resonance because the ground state deep inside these regions are BCS superfluid and Bose-Einstein condensate (BEC) of molecules respectively.  The intriguing region is near resonance ($\delta \rightarrow 0$) where the system becomes strongly interacting.  The experiments in ref.\cite{Jin} and \cite{Ketterle} explore this region.

In ref.\cite{Jin}, an equilibrium state was prepared at an initial field $B_{i}$, which was quickly changed to $B_{f}$ 
deeper in the BEC side. It is found that when $B_{i}$ is sufficiently close to resonance, a molecular condensate will emerge at $B_{f}$ such that the fraction of condensed molecules ($N_{o}/N_{m}$) is independent of the (fast) sweep rate\cite{Jin}. 
The JILA group considers this as evidence for condensation of fermion pairs in the initial state. Their picture is the BCS-BEC crossover theory, which describes the BCS to BEC evolution as a smooth reduction in size of the condensed pair\cite{single,Randeria,two}.
The immediate question is that near resonance the size of the pair should be of
interparticle spacing, then the direct overlap between these pairs and the tightly bound molecules at $B_{f}$ should be very small.   In that case, the final molecular condensate must be generated by the fast sweeping field, and the relation between the momentum distributions in the final state and the initial state is not immediately clear. 

In contrast, the MIT group thinks that their similar results are due to a BEC of tightly bound molecules in closed channel persisting onto the BCS side. They argue that even though these molecules are unstable in vacuum on that side, they can be stabilized by the Fermi sea.  Somewhat earlier,  Falco and Stoof (FS)\cite{FS} had also suggested that the molecular fraction $N_{0}/N_{m}$ observed at $B_{f}$ in ref.\cite{Jin} is given by the molecular component $Z(\delta)$ 
in the condensed fermion pair in the initial state, $|{\rm pair}\rangle=\sqrt{Z(\delta)} |{\rm closed}\rangle + \sqrt{1-Z(\delta)}|{\rm open}\rangle$, where $|{\rm closed}\rangle$ and $|{\rm open}\rangle$ are closed channel and open channel fermion pairs. Using a phenomenological resonance model, FS calculated the boundary of vanishing $Z$ and found reasonable match
with the boundary of vanishing $N_{0}/N_{m}$ observed in ref.\cite{Jin}.  FS argue that even though when $Z$ vanishes, the system still has pairing in open channel, and the observed boundary of vanishing $N_{0}/N_{m}$ is a sharp crossover from molecular rich pair to fermion pair, (see fig.1 in ref.\cite{FS}).  This ``persisting molecular BEC" pictures\cite{Ketterle,FS} are very different from the usual crossover picture, which does not predict a sharp division between molecule rich and pure fermion pairs. 

Adding to this confusion is the question of whether the many-body physics near a Feshbach resonance can only be described by 
``two channel" (or ``resonance") models which incorporate open and closed channel physics specifically\cite{two}, or can be 
described by simpler ``single channel" models which consist of two types of fermions interacting with a tunable potential\cite{single,Randeria}.  Should these models have very different properties 
at resonance,  and if the physics near resonance is universal as generally believed, it would imply that there are two distinct universality classes,  and that only the two channel universality class 
gives the right physics of Feshbach resonance.  The other possibility is that two channel models are not universal, and that the notion of universality inferred from current experiments\cite{Duke,ENSenergy} is misconstrued. 
 
In this paper, we point out that these confusions can be eliminated by deriving the precise expression for the observed molecular fraction, calculating the properties of the system explicitly using crossover theory, and examining the nature of the broken symmetry. We first perform $T=0$ calculations using crossover theory\cite{single,Randeria} for three different types of potential: $V_{I}$: a square well, $V_{II}$: a well with a high barrier, and $V_{III}$ a delta-function potential.  Since $V_{II}$ can accommodate a very long lived quasi-bound state, it has the major characteristics of a two channel (or  ``resonance") model. We shall show that all these potentials give rise to identical 
properties (such as  chemical potential, energy gap, and coherence factors) near resonance, demonstrating universality in this regime. 
The fact that $V_{II}$ carries all the characteristics of the two-channel model suggests both types of models may have identical properties near resonance.  To our surprise, our calculations show that there $is$ non-trivial overlap between the pairs at resonance (denoted as  $x=0$, where $x=(k_{F}a_{s})^{-1}$
and $k_{F}$ is the Fermi wavevector)  and those at $x$ as large as 12,  which corresponds to the the final field $B_{f}$ in ref.\cite{Jin}.  This eliminates a serious concern formerly expressed by one of us (TLH)\cite{Physics} about the validity of  projection argument in ref.\cite{Jin}.  
To the extent that direct projection describes the experimental process, one can show that $N_{0}/N_{m}$ is directly proportional to the superfluid order parameter, and 
 the boundary for vanishing $N_{0}/N_{m}$ in $T-B$ plane observed in ref.\cite{Jin,Ketterle} is precisely the $T_{c}$ between superfluid and normal gas, {\em not} a crossover between different types of pairs.  Since $V_{I}$ to $V_{III}$ have identical properties near resonance, one can use the simplest potential ($V_{III}$) to study the finite temperature properties in this region. Such study 
 had been performed in ref.\cite{Randeria} by studying Gaussian fluctuations about the mean field state. Remarkably, the phase boundary predicted in ref.\cite{Randeria} matches  very well with the observed boundary for vanishing condensate fraction ($N_{0}/N_{m}=0$)\cite{Jin,Ketterle}  with no adjustable parameters.

{\bf (A)  The fraction of condensed molecules:} 
Consider a fermion pair with total momentum ${\bf q}$, 
$D^{\dagger}_{\bf q}(x) = \sum_{{\bf k}, \alpha\beta} f^{}_{{\bf k}, \alpha\beta}(x) a^{\dagger}_{{\bf k}+{\bf q}/2, \alpha}
a^{\dagger}_{-{\bf k}+{\bf q}/2, \beta} /2$, where $a^{\dagger}_{{\bf k},\alpha}$ creates a fermion with momentum ${\bf k}$ and hyperfine spin $\alpha$, $x= (k_{F}a_{s})^{-1}$
describes the distance from resonance, 
and $f^{}_{{\bf k}, \alpha\beta}(x)$ is the Fourier transform of the pair wavefunction 
$f_{\alpha\beta}({\bf r}; x) = \Omega^{-1/2} \sum_{\bf k} e^{i{\bf k}\cdot {\bf r}} f^{}_{{\bf k}, \alpha\beta}(x)$ which depends on $x$; $\Omega$ is the volume of the system. 
A condensate of $N$ zero momentum pairs  is $|x \rangle = {\cal N}D^{\dagger N}_{{\bf q =0}}(x) |{\rm vac}\rangle$.
It  reduces to a weak coupling BCS superfluid and a molecular condensate as $x<<-1$ and $x>>1$ respectively\cite{single,Randeria}.  In grand canonical description,   this pairing state becomes 
\begin{equation}
|\Psi(x)\rangle = {\cal N}\prod_{{\bf k}, \alpha\beta} \left( u_{\bf k}^{}(x)+ v_{{\bf k}, \alpha\beta}^{}
(x) a^{\dagger}_{{\bf k}, \alpha}a^{\dagger}_{-{\bf k}, \beta}\right)|{\rm vac}\rangle
\label{BEC}\end{equation}
where ${\cal N}$ is the normalization constant, $u$ and $v$ are related to the Fourier transform of 
$f_{\alpha\beta}(r) $ as $f_{{\bf k}, \alpha\beta} = \zeta v_{{\bf k}, \alpha\beta}/u_{\bf k}$, and $\zeta$ is a normalization constant. 

Let $x_{o}$ and $x$ be the distances of the initial and final state from resonance, and $x$ be on the BEC side. 
When the system is jumped from $x_{o}$ to $x$, the number of condensed and un-condensed molecules emerge at $x$ are
$N_{0} = \langle  D^{\dagger}_{\bf 0}(x)D^{}_{\bf 0}(x)    \rangle_{x_{o}} $
and $N_{ex} =\sum_{{\bf q}\neq 0} \langle  D^{\dagger}_{\bf q}(x)D^{}_{\bf q}(x)    \rangle_{x_{o}}$ respectively.  We then have from eq.(\ref{BEC}), 
\begin{equation}
N_{0} = |\langle D_{\bf 0}(x)\rangle|^2 =\left|  \sum_{\bf k} f_{{\bf k},\alpha\beta}(x)
\Psi^{\ast}_{{\bf k}, \alpha\beta}(x_{o})/2\right|^{2}, 
\label{N0} \end{equation}
$N_{ex}$$=$$\sum_{\bf k, q}$$|f_{{\bf k},\alpha\beta}(x)|^2$$n_{{\bf q}, \alpha}$$n_{{\bf q}-2{\bf k}, \beta}/2$, 
where $\Psi^{\ast}_{{\bf k}, \alpha\beta}(x_{o}) = \langle a^{\dagger}_{{\bf k}, \alpha} 
a^{\dagger}_{-{\bf k}, \beta}\rangle_{x_{o}}$, $n_{\bf k} = \langle a^{\dagger}_{{\bf k}, \alpha}
 a^{}_{{\bf k}, \alpha}\rangle_{x_{o}}$, and we have ignored in eq.(\ref{N0}) a term down by a factor of $N^{-1}$.  {\em Eq.(\ref{N0}) shows that $N_{0}$ is the overlap of the initial order parameter $\Psi_{\alpha\beta}$  with the final pair wavefunction.}   

Let us first discuss the single channel case because it is illuminating. 
Denoting the two spin states as $\uparrow$ and $\downarrow$, the s-wave pairing function is 
$f_{\alpha\beta}(r) = f(r) i\sigma^{y}_{\alpha\beta}$,  
and 
\begin{equation}
\frac{N_{0}}{N_{m}} = \frac{| \sum_{\bf k} f_{\bf k}(x) \Psi^{\ast}_{\bf k} |^2}
{| \sum_{\bf k} f_{\bf k}(x) \Psi^{\ast}_{\bf k} |^2 + \sum_{{\bf k}, {\bf q}} 
| f_{\bf k}(x) |^2 n_{\bf q}n_{{\bf q}- 2{\bf k}} }.
\label{fraction} \end{equation}
where $N_{m}= N_{0}+ N_{ex}$, $\Psi({\bf k})^{\ast} = \langle a^{\dagger}_{{\bf k}, \uparrow} a^{\dagger}_{-{\bf k}, \downarrow}\rangle_{x_{o}}$, $n_{\bf k}= \langle a^{\dagger}_{{\bf k}, \uparrow} a^{}_{{\bf k}, \uparrow}\rangle_{x_{o}} = \langle a^{\dagger}_{{\bf k}, \downarrow} a^{}_{{\bf k}, \downarrow}\rangle_{x_{o}}$. Since $\Psi_{\bf k}$ vanishes at $T=T_{c}$, {\em the  curve $N_{0}/N_{m}=0$ in the $T-B$ plane is therefore the boundary of superfluid to normal transition}.  

{\bf (B) The  Crossover Picture:} The basic assumption of crossover theory is that the ground state near resonance is still given by eq.(\ref{BEC}). We have studied eq.(\ref{fraction}) using the single channel  crossover theory for: {\bf (I)} a square well,  $V_{I}(r) = -|U_{o}|$ (or $0$)  for $r<r_{o}$ (or $r>r_{o}$); {\bf (II)} a well plus a barrier, $V_{II}(r) =  -|U_{o}|$ for $r<r_{o}$. 
 $V_{II}(r) =  |U_{1}|$ for $r_{o}<r<r_{1}$,   and $V_{II}(r)=0$ for $r>r_{1}$; 
 {\bf (III)} a contact potential $V_{III}(r) = g\delta(r)$, where $g= 2\pi \hbar^2 a_{s}/M$, where $M$ is the mass of the fermion.  The Hamiltonian is 
$ H = \sum_{{\bf k}, \alpha}  \epsilon_{\bf k} a^{\dagger}_{{\bf k},\alpha}a^{}_{{\bf k},\alpha} 
 + \sum_{{\bf k}, {\bf k'}, {\bf q}}^{} V({\bf k}-{\bf k'}) a^{\dagger}_{{\bf q}/2 + {\bf k}, \uparrow} 
 a^{\dagger} _{{\bf q}/2 - {\bf k}, \downarrow}  a^{}_{{\bf q}/2 - {\bf k'}, \downarrow}
 a^{}_{{\bf q}/2 + {\bf k'}, \downarrow}$, $\epsilon_{\bf k} = \hbar^2k^2/2M$. 
The range of the potential of $V_{I}$ and $V_{II}$ ($r_{o}$ and $r_{1}$) are taken to be of atomic size, much smaller than interparticle spacing, i.e. $k_{F}^{} r_{o}, k^{}_{F} r_{1}<< 1$.  By varying $|U_{o}|$, a bound state can be peeled off from the continuum, causing $a_{s}$ to jump from $-\infty$ to $+\infty$.  The main difference between $V_{I}$ and $V_{II}$ is that the latter can accommodate a quasi-bound state of atomic size, similar to the feature of the two channel or the resonance model. 

For all cases, $u_{\bf k}$ and $v_{\bf}$ are given by the well known expressions
$|u_{\bf k}|^2 = (E_{\bf k}^{} + \xi^{}_{\bf k})/(2E^{}_{\bf k})$, $
|v_{\bf k}|^2 = (E_{\bf k}^{} - \xi^{}_{\bf k})/(2E^{}_{\bf k})$, 
where $\xi^{}_{\bf k} = \epsilon^{}_{\bf k} - \mu$, $E^{}_{\bf k} = 
\sqrt{\xi^{2}_{\bf k} + \Delta^{2}_{\bf k}}$, and the energy gap  $\Delta^{}_{\bf k}$ at $T=0$ is given by the gap equation 
$\Delta^{}_{\bf k} = - \sum_{\bf k'}^{} V({\bf k}- {\bf k'})  \Delta^{}_{\bf k'}/(2E_{\bf k'}^{})$.
The corresponding expressions for $\Psi_{\bf k}^{}$ and $n_{\bf k}$ are $\Psi^{}_{\bf k} = u_{\bf k}^{} v^{}_{\bf k}= \Delta^{}_{\bf k}/(2E_{\bf k})$, $n_{\bf k} = v^{2}_{\bf k}$. 
The chemical potential $\mu$ is determined by the number density $n$ as 
$n =\Omega^{-1} \sum_{\bf k} n_{\bf k}^{}$$=n(T=0, \mu)$. 

\begin{figure*}
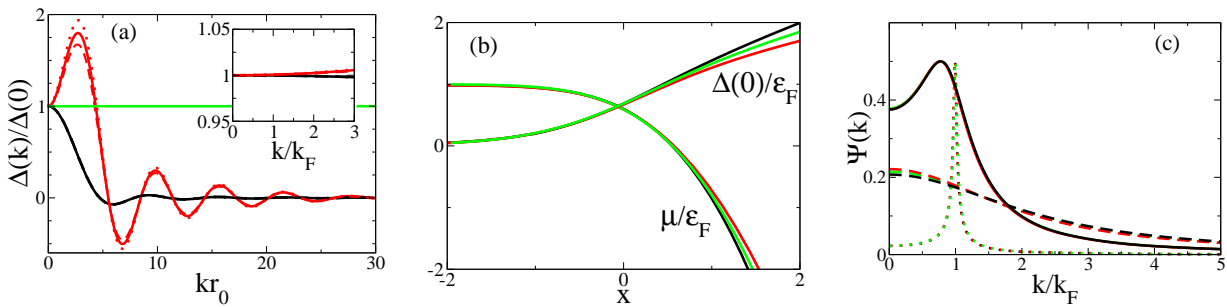

$\begin{array}{c@{\hspace{0.2in}}c@{\hspace{0.2in}}c}
\epsfxsize=2in
\epsffile{Fig1_a.eps} &
\epsfxsize=2in
\epsffile{Fig1_b.eps} &
        \epsfxsize=2in
        \epsffile{Fig1_c.eps} 
        \end{array}$
\caption{Fig.1a to 1c:  
Black, red, and green represent $V_{I}, V_{II}, V_{III}$, $x=(k_{F}a_{s})^{-1}$, and $k_{F}r_{o}= 0.05$.  
Dashed, solid, and dotted lines denote $x=2, 0, -2$. Within the range $-2<x<2$, there are little changes in $\Delta(k)/\Delta(0)$, although $\Delta(0)$ and $\mu$ undergo significant changes (see fig.1b), leading to very different $\Psi(k)$ in fig. 1c. }
\label{figtest-fig}
\end{figure*}

{\bf (C) Results:}  We have solved the gap equation numerically and have inverted the relation $n=n(\mu)$ so that quantities like
$\Delta_{\bf k}^{}, u^{}_{\bf k}, v^{}_{\bf k}, \Psi_{\bf k}^{}, n^{}_{\bf k}$ are functions of the parameters of the potential (denoted collectively as $\{ Y_{i} \}$) and density $n$.  Our results are displayed in figure 1 to 9.  In these figures, 
we use the dimensionaless parameter $x= (k^{}_{F}a_{s})^{-1}$ instead of potential parameters $\{ Y_{i}\}$. The results for potential 
$V_{I}$ ,$V_{II}$, and $V_{III}$ will be colored in black, red, and green respectively.  
For both $V_{I}$ and $V_{II}$, we take $k_{F}r_{o}= 0.05$. For $V_{II}$, we choose $r_{1}/r_{o}=1.2$ and $U_{1}=  
5|U_{0}^{(min)}|$, where $|U_{0}^{(min)}|= \hbar^2\pi^2/(4Mr^{2}_{o})$ is the value for a square well to produce a bound state.

{\bf (C1)  $\Delta_{\bf k}$, $\Delta_{o}$, and $\mu$:}  In figure 1a, we have plotted $\Delta_{\bf k}/\Delta_{0}$ versus $kr_{o}$ for $x=2,0, -2$, denoted as dashed, solid, and dotted lines respectively.  The inset shows that  $\Delta_{\bf k}/\Delta_{0}$  for different potentials have similar values at low momenta (even up to a few $k_{F}$'s) but differ significantly at higher momenta, though it vanishes for realistic potentials 
$V_{I}$ and $V_{II}$.  Within the range $-2<x<2$, this ratio is essentially unchanged.  The value of $\Delta(0)$ and $\mu$, however, changes a lot as $x$ varies from -2 to 2, as shown in fig.1b.  These variations lead to the large 
differences in $\Psi_{\bf k}$ as one move across the resonance. It is clear from fig.1b and 1c that universal behavior sets in around 
resonance ($x=0$) since different potentials give identical results. 
For all potentials, we find that $n_{k}$ is accurately described by that of 
$V_{III}$.  Its figure is given in ref.\cite{Randeria}. 

\begin{figure*}
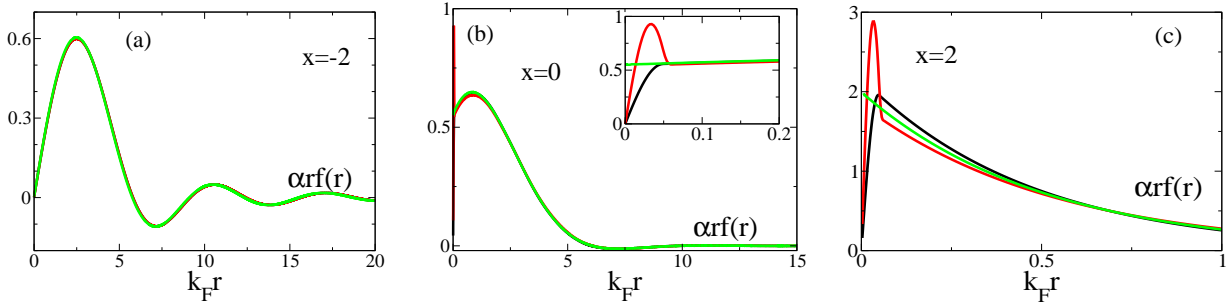

$\begin{array}{c@{\hspace{0.2in}}c@{\hspace{0.2in}}c}
\epsfxsize=2in
\epsffile{Fig2_a.eps} &
\epsfxsize=2in
\epsffile{Fig2_b.eps} &
        \epsfxsize=2in
        \epsfysize=1.57in 
        \epsffile{Fig2_c.eps} 
        \end{array}$
\caption{Fig.2a to 2c: Pair wavefunctions for different values of $x$, $\alpha=\sqrt{4\pi/k_F}$.
For $V_{II}$, $f$ has a peak at $r<r_{o}$, yet most of the weight for all potentials is outside the atomic range $r_{o}$.  At resonance the size of $f$ is given by $k_{F}^{-1}$. }
\end{figure*}

\begin{figure*}
$\begin{array}{c@{\hspace{0.2in}}c@{\hspace{0.2in}}c}
\epsfxsize=2in
\epsffile{Fig3_a.eps} &
\epsfxsize=2in
\epsfysize=1.55in 
\epsffile{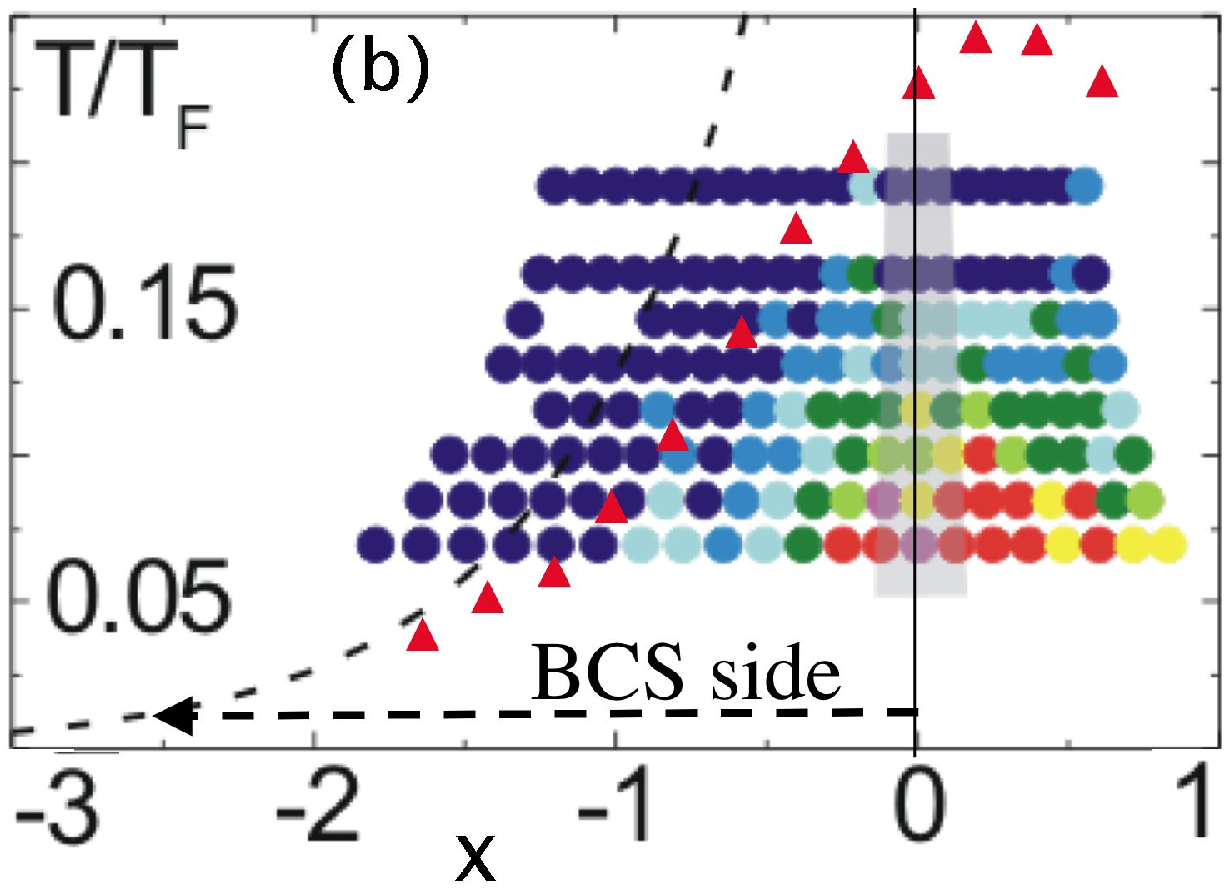} &
\epsfxsize=2in
        \epsffile{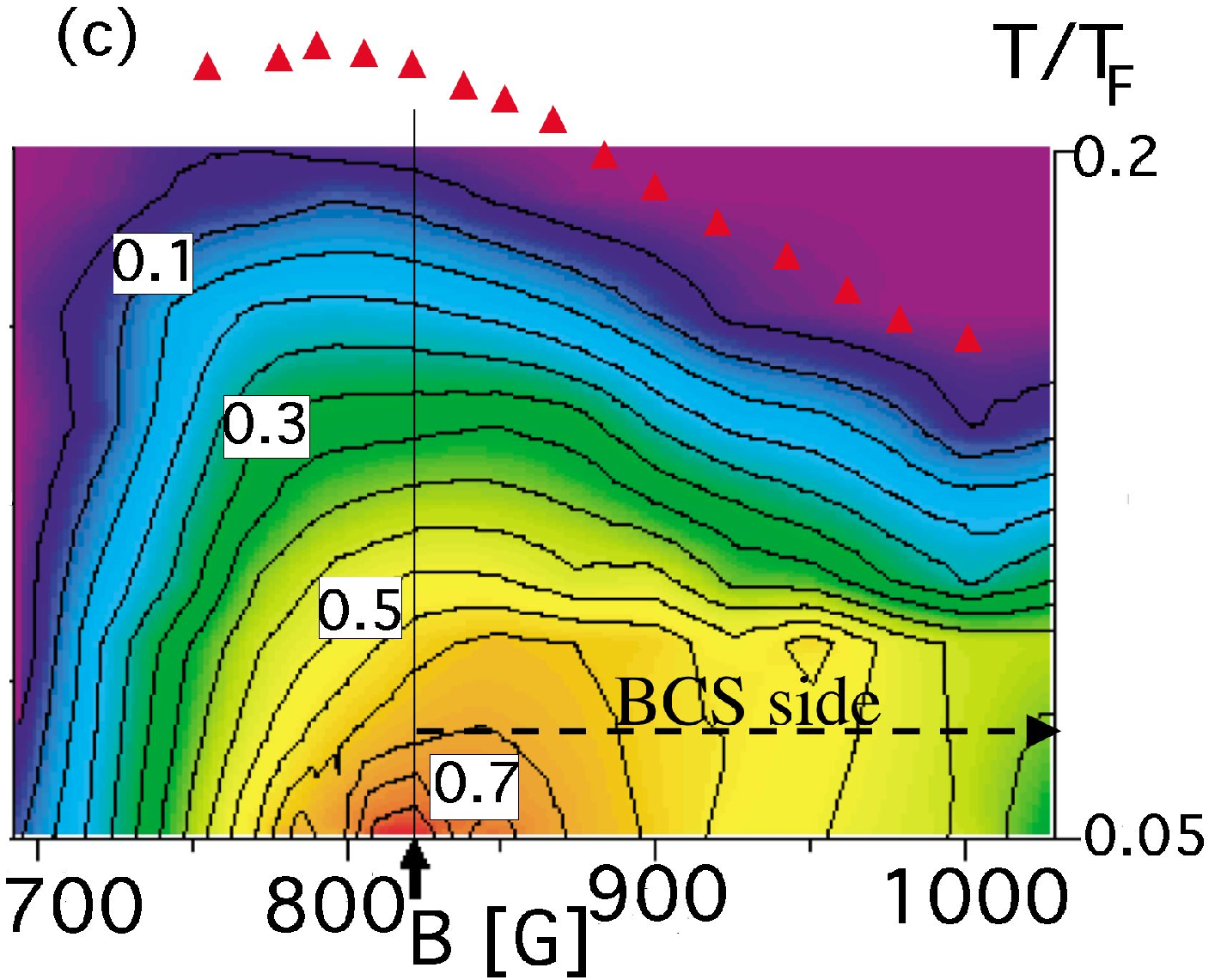} 
        \end{array}$
\caption{Fig.3a: The $T=0$ value of $N_{0}/N_{m}$ at $x=12$ as a function of initial state $x_{o}$  is represented by the solid line. The data of ref.\cite{Jin} and \cite{Ketterle} are denoted  as circles and squares.  As the final state $x$ moves toward the BEC side, the curves moves toward the BCS side, reaching a fixed asymptote, which depends on the microscopic nature of $\Delta(k)$ at large momenta (see fig.1a). The asymptotes of  $V_{I}$ and $V_{II}$ are represented by the dotted and dashed curve resp.  In fig.3b and 3c, we superpose the phase boundary calculated in ref.\cite{Randeria} (red triangle) on the data of ref.\cite{website} and \cite{Ketterle}.  In fig.3b, light color means large $N_{0}/N_{m}$.  
 The numeral correspondence of the color code is given in ref.\cite{website}.  Dark blue corresponds to $N_{0}/N_{m}=0$.
In fig.3c, we have used the approximate formula $x=(832-B)/(B-640)$ to convert magnetic field into $x=(k_{F}a_{s})^{-1}$\cite{estimate}.} 
\end{figure*}

{\bf (C2) Pair wavefunction $f$: }  To display $f$, we plot $\sqrt{4\pi/k_{F}}rf(r)$ versus $k_{F}r$ in fig.2a to 2c. 
One sees that all potentials have essentially the same pair wavefunctions unless one goes deeper into the BEC side.  The function $f$ on the BEC side for all 
potentials have essentially the Feshbach form $\propto e^{-r/a_{s}}/r$. Until $a_{s}$ reduces to the atomic size $r_{o}$, $f$ has a non-trivial fraction beyond $r_{o}$. It can therefore have significant 
overlap with the pair wavefunction near resonance.  In fact, the overlap between $f$ at $x=0$ and that at $x=12$ (which is the final state in ref.\cite{Jin}) is as big as $18 \%$.  
Thus, the direct projection picture suggested in ref.\cite{Jin} (formerly criticized by one of us\cite{Physics}) is in fact reasonable. 
In the case of $V_{II}$, we note that while a molecular component of range $r_{o}$ is visible on the BCS side near resonance, it essentially vanishes for $x<-2$. However, the $T=0$ condensate fraction $N_{0}/N_{m}$ is still very large for initial 
 state $x_{o}=-3$ which has no molecular component.  {\em This shows that the molecular component is unimportant in contributing to the projected condensate fraction $N_{0}/N_{m}$}.  

{\bf (C3) The condensate fraction $N_{0}/N_{m}$, and the superfluid-normal phase boundary:} In figure 3a, we show the $T=0$ molecular fraction $N_{0}/N_{m}$ at $x=12$ as a function of the initial state $x_{o}$.  
The data of ref.\cite{Jin} and \cite{Ketterle} are represented by circles and square respectively.  The former is obtained at $T/T_{F}=0.05$. 
The temperature of the latter is uncertain.  Since the decrease of $N_{0}/N_{m}$ on the BEC side
 appears to be related to three body effects\cite{Jin},  we apply equilibrium theories only to the BCS side ($x<0$).  The large difference in scale between the results at $T=0$ and at $T/T_{F}=0.05$ is anticipated since $T_{c}$ is strongly suppressed in the strongly interacting regime\cite{single,Randeria}, which will reflect in a strong suppression of $N_{0}/N_{m}$.
The full effect of the crossover theory is best illustrated by showing $T_{c}$ in the $T-x$ plane.  In fig.3b and 3c, we have 
superposed the $T_{c}$ predicted in ref.\cite{Randeria} on the contour plots for $N_{0}/N_{m}$ in the JILA\cite{website} and the MIT\cite{Ketterle} experiments. (Because of  the universality near resonance demonstrated in fig.1 and 2, we can use the finite temperature results for $V_{III}$\cite{Randeria} to compare with experiments.) 
 The matches are remarkable, in view that there are no adjustable parameters. 

In fig.3a, we also show that as the final state moves deeper into the BEC side, the boundary of vanishing $N_{0}/N_{m}$ moves to the opposite, i.e. BCS side. This is because while both $N_{0}$ and $N_{ex}$ decrease as the final state becomes more tightly bound, the decrease of $N_{ex}$ is faster than that of $N_{0}$ so that $N_{0}/N_{m}$ actually increases. This fact can be demonstrated numerically, but can be established analytically if we take $f(k) = (k^2 + \Lambda^2)^{-1}$ and study the change of $N_{0}/N_{ex}$ as a function of $\Lambda^2$. 
As $\Lambda\rightarrow \infty$,  $N_{0}/N_{m}$  reaches the asymptotic curve $\Gamma$: 
$N_{0}/N_{m} =A/(A+B)$, where $A= | \sum_{\bf k}\Psi_{\bf k}^{}|^2$, and 
$B=\sum_{\bf q, k} n_{\bf k}^{} n^{}_{{\bf k}- {\bf q}}$, a prediction that can be verified experimentally. 
Since jumping to fields where molecular size are comparable $r_{o}$ (such as a deeply bound state) means sampling short distance behaviors of the order parameter (see fig.1a), $\Gamma$ will not be universal, as illustrated in the dashed and dotted curve in fig.3a.  

{\bf (D) Two channel models:}  For Feshbach resonances, $f_{\alpha\beta}$ in Section {\bf (A)} is  $f_{\alpha\beta}(r)$$=$$\sqrt{Z}$$f^{(c)}(r)\chi^{(c)}_{\alpha\beta}(r)$ $+$$\sqrt{1-Z}$$ f^{(o)}(r)\chi^{(o)}_{\alpha\beta}(r)$, where $f$ and $\chi$ are normalized orbital and spin functions, $``c"$ and $``o"$  denote close and open channel, and $\chi^{(c)}$ and $\chi^{(o)}$ are orthogonal.  
The zero momentum pair is $D^{\dagger}_{\bf q=0} = \sqrt{Z} b^{\dagger}_{\bf q=0} + \sqrt{1-Z}
\sum_{\bf k} f^{(o)}_{\bf k}\chi^{(o)}_{\alpha\beta}a^{\dagger}_{{\bf k}\alpha}a^{\dagger}_{ -{\bf k}, \beta}$, where $b^{\dagger}_{\bf q =0} = \sum_{\bf k} f^{(c)}_{\bf k}\chi^{(c)}_{\alpha\beta}a^{\dagger}_{{\bf k}\alpha}a^{\dagger}_{ -{\bf k}, \beta}$ is the closed channel boson in the 
resonance model\cite{two}.  From eq.(\ref{N0}), we have $N_{0}= |\sqrt{Z}\sum_{\bf k} f^{(c)}_{\bf k}\Psi^{(c)\ast}_{\bf k} + \sqrt{1-Z}\sum_{\bf k} f^{(o)}_{\bf k}\Psi^{(o)\ast}_{\bf k}|^2$, where $\Psi_{\bf k}^{(c,o)} = {\rm Tr} \chi^{(c,o) \dagger} \Psi_{\bf k}/2$ are the components of the order parameter $\Psi_{ {\bf k}, \alpha\beta}$ in closed and open channel.
If the final state is a tightly bound molecule ($Z=1$), then the vanishing of $N_{o}$ is given by the vanishing of $\Psi^{(c)}$.  However, to infer from a vanishing $\Psi^{(c)}$ of a possible non-zero $\Psi^{(o)}$, (like the vanishing $Z$ in $f$ means the open channel component $(1-Z)\rightarrow 1$), is incorrect. This is because $\Psi_{{\bf k},\alpha\beta}$ has no normalization constraint,  and
both $\Psi^{(c)}_{\bf k}$ and $\Psi^{(o)}_{\bf k}$ are dynamically connected through the coupling term $b^{\dagger}_{\bf q=0}a^{}_{{\bf k}\alpha} a^{}_{-{\bf k}\beta}\chi^{(o) \dagger}_{\beta\alpha} + h.c.$ in the Hamiltonian. Any non-vanishing pairing order in one channel will generate a non-vanishing order  in  the other.    

We have also learned from D. Jin that for $^{40}$K there is a shallow bound state (${\cal C}$) in the open channel. 
Thus, as the close channel bound state (${\cal B}$) is Zeeman shifted down, it will cross the zero energy state (${\cal A}$)  and then the bound state (${\cal C}$)  in the open channel. 
One can then go from ${\cal A}$ to ${\cal B}$ to ${\cal C}$ using a magnetic field sweep, so that  a pair of fermions in the open channel can be swept into a bound state in the same channel. 
In that case, the system can be described by single channel models.  

This work is supported by NASA GRANT-NAG8-1765  and NSF Grant DMR-0109255.

\end{document}